\documentclass{eas}
\usepackage{graphicx}
\usepackage{epsfig,color,multicol}
\newcommand{\indx}[1]{ \textrm{\scriptsize #1} }
\usepackage{astron}
\begin{document}
\bibliographystyle{astron}

\title{PIERNIK MHD code --- a multi--fluid, non--ideal\\
 extension of the relaxing--TVD scheme~(III)} 
\runningtitle{M. Hanasz \etal : PIERNIK MHD code \dots ~(III)}
\author{Micha\l{} Hanasz}
\address{Toru\'n Centre for Astronomy, Nicolaus Copernicus University, Toru\'n, Poland; 
\\ \email mhanasz@astri.uni.torun.pl}
\author{Kacper Kowalik}\sameaddress{1} 
\author{Dominik W\'olta\'nski}\sameaddress{1} 
\author{Rafa\l{} Paw\l{}aszek}\sameaddress{1} 
\begin{abstract}

We present a new multi--fluid, grid MHD code PIERNIK, which is based on the
Relaxing TVD scheme~\cite{Jin95}. The original scheme (see Trac \& Pen~\cite*{trac} and Pen~\etal~\cite*{pen}) has been extended by an addition of
dynamically independent, but interacting fluids: dust and a diffusive cosmic ray
gas, described within the fluid approximation, with an option to add other
fluids in an easy way.  The code has been equipped with shearing--box boundary
conditions, and a selfgravity module, Ohmic resistivity module, as well as other
facilities which are useful in astrophysical fluid--dynamical simulations. The
code is parallelized by means of the MPI library. In this paper we 
present Ohmic resistivity extension of the original Relaxing TVD MHD scheme, and
show examples of magnetic reconnection in cases of uniform and
current--dependent resistivity prescriptions.

\end{abstract}
\maketitle
%
\section{Resistive MHD equations}
\enlargethispage{\baselineskip}
Piernik MHD code is capable of dealing with resistive MHD equations \linebreak
\cite{pawlaszek-08},
which involve the resistive dissipation term in the induction equation
\begin{equation}
\label{induction_equation}
\partial_t \mathbf{B}
= \nabla\times \left(\mathbf{v}\times\mathbf{B} \right)-\nabla\times(\eta\mathbf{J}),
\end{equation}
where $\mathbf{J}$ is the electric current density. Due to the conservative
nature of the basic MHD algorithm, any decrement of magnetic energy during the
update of magnetic field results in a corresponding increment of thermal
energy.  Therefore, no additional source term related to resistivity is needed
in the energy equation.
\par The resistivity algorithm of PIERNIK code relies on additional terms of
resistive origin in the electromotive force, supplemented to the original
Constraint Transport (CT) scheme implemented by Pen~\etal~\cite*{pen} within the
Relaxing TVD scheme. In the present version of PIERNIK we have implemented a 
current--dependent resistivity, according to Ugai~\cite*{ugai}
\begin{equation}
\eta=\eta_{1}+\eta_{2}(J^2-J^2_{\indx{crit}})\Theta(|\mathbf{J}|-|\mathbf{J}_{\indx{crit}}|),
\end{equation}
where $\eta_{1}$ is a constant corresponding to uniform resistivity, $\eta_{2}$
represents the anomalous resistivity, which switches on when the current density
exceeds a certain critical value $J_{\indx{crit}}$,  and $\Theta$ is the
Heaviside step function. 
\section{Resistivity algorithm}
\begin{figure}[!ht]
\centering
\includegraphics[width=0.39\textwidth]{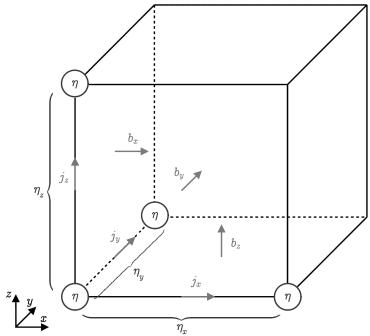}
\caption{Position of the magnetic field $\mathbf{B}$ components, the electric current $\mathbf{J}$ components and the magnetic resistivity $\eta$ on a computational cell. Braces indicate interpolation of $\eta$ to the position of electric current density $\mathbf{J}$} 
\label{siatka}
\end{figure}

The relation between electric currents and magnetic fields in magnetized media 
is given by  Ampere's law
\begin{equation}
\mathbf{J}=\nabla\times\mathbf{B},
\end{equation}
where the code normalization of current densities  is applied, to get rid of the
factor $c/4\pi$ in Ampere's law. To solve the Ampere's  equation numerically in
a grid scheme  we set the location of current components at the cell edges, while
the current--dependent resistivity values are initially placed at the cell corners
(see Fig.~\ref{siatka} for details). 
To improve the stability of the resistive part of the MHD scheme a weighted
average of resistivity among the closest neighbours is computed before resistive terms are taken into account in the computation of electromotive forces.
Resistivity derived in this fashion is then interpolated to the positions of 
the current density components (cell edges) (see Fig.~\ref{siatka}) and then the resistive part of the electromotive
force $({\cal E}_\eta =\eta\mathbf{J})$ is computed.
Eventually, we compute the diffusive terms of induction equation (\ref{induction_equation}) and update magnetic field components. 
\section{Current sheet test}
The following  2D tearing instability setup has been proposed by Hawley \& Stone
\cite*{hawley-stone-95} (see also the web--page of ATHENA code~\cite{athena}) to test robustness of
MHD algorithms at extreme conditions. The initial setup assumes  magnetic field
$B_y = \pm 1$, which changes sign at two locations $(x = \pm 0.25)$, in a domain
$-0.5 \leq x \leq 0.5$ and $-0.5 \leq y \leq 0.5$ with doubly periodic boundary
conditions.  Initial density and pressure are uniform: $\rho = 1$, $p=\beta/2$,
where $\beta=0.1$ is the 'plasma--$\beta$'. The initial sinusoidal velocity
perturbation is perpendicular to magnetic field.  The initial setup consists of
two current sheets at $x = \pm 0.25$, which promote magnetic reconnection,  due to the numerical or physical resistivity. 
\par
\begin{figure}[!ht]
\centering
\includegraphics[width=0.30\columnwidth]{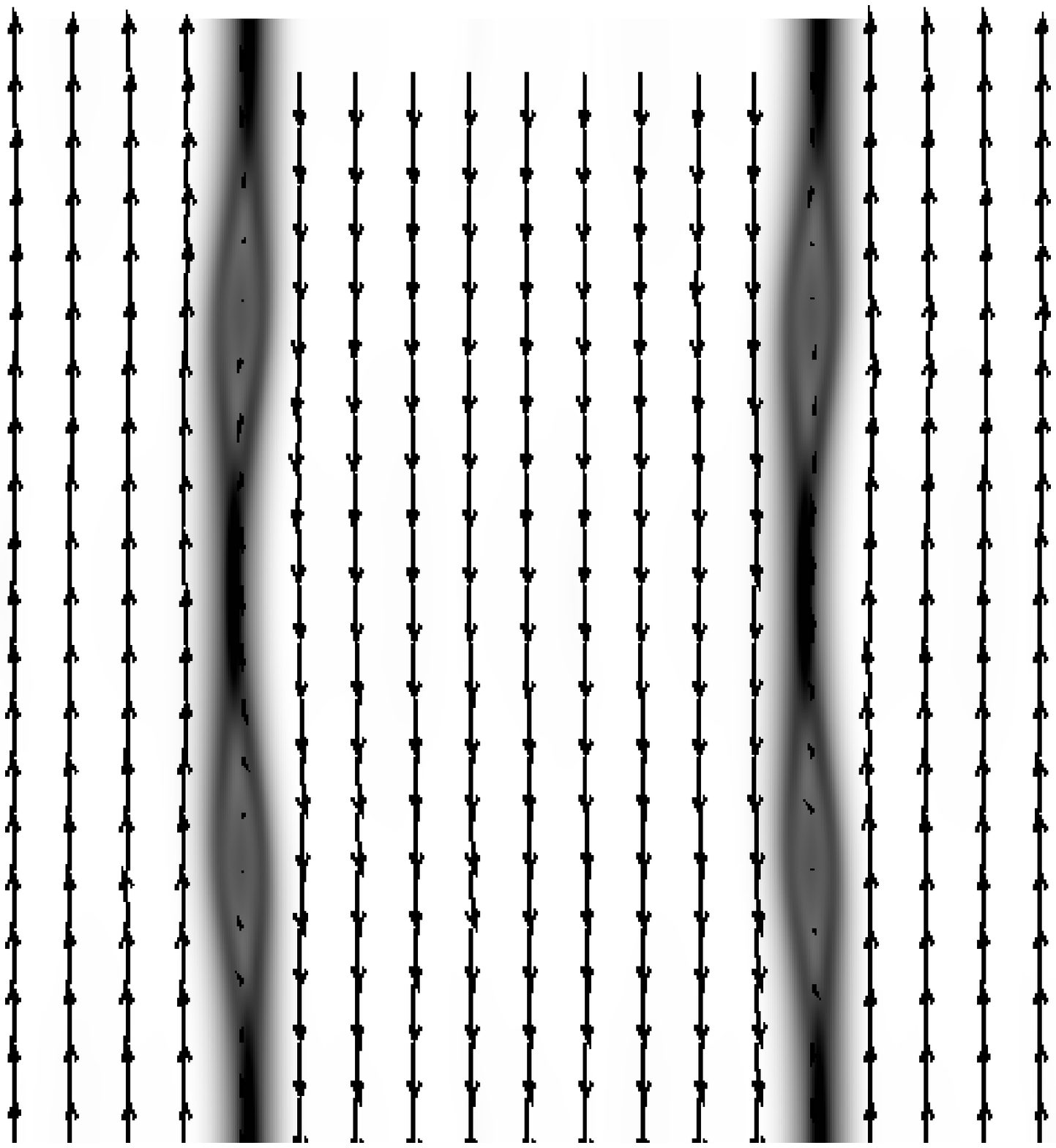}
\includegraphics[width=0.30\columnwidth]{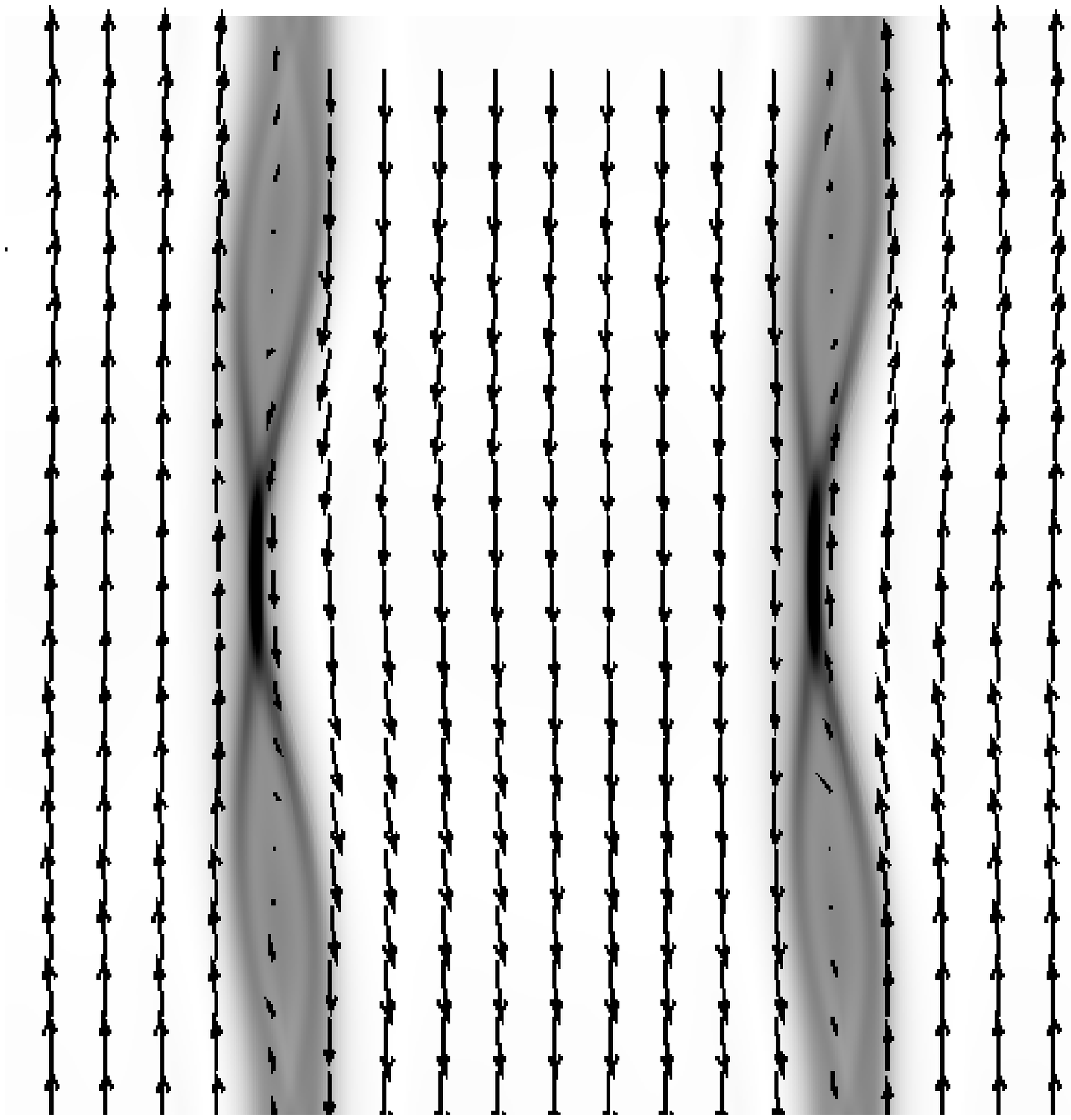}
\includegraphics[width=0.30\columnwidth]{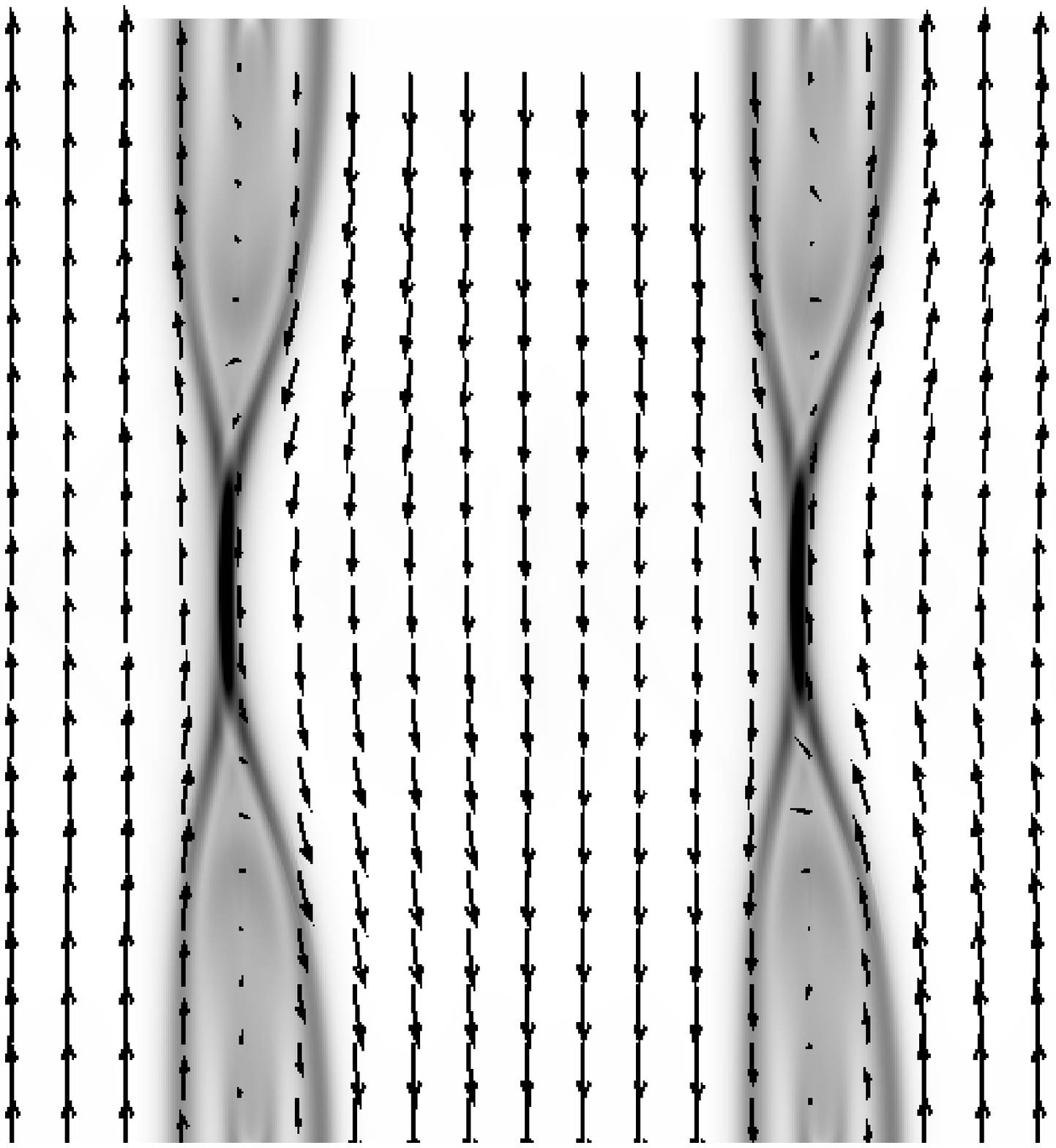}
\caption{Snapshots of tearing instability at $t=2, 4, 6$ with an uniform
resistivity. Grayscale represents current density and arrows denote normalized
magnetic field. Resolution $512^2$}
\label{fig:uniform-resistivity}
\end{figure}

 We shall use the above setup to demonstrate qualitative differences between
simulations with the uniform and current--dependent resistivity prescriptions.
For the uniform resistivity test we assume $\eta_1=10^{-4}$ and $\eta_2=0$.  It
is apparent that current sheets form elongated structures, that look rather
smooth, since the uniform resistivity acts everywhere in the computational
domain. The uniform resistivity prescription applied in this experiment
(Fig.~\ref{fig:uniform-resistivity}), leads to configurations resembling
the Sweet--Parker model (Parker \cite*{parkerrec}; Sweet \cite*{sweetrec}). 
\enlargethispage{\baselineskip}
\par
\begin{figure}[!ht]
\centering
\includegraphics[width=0.30\columnwidth]{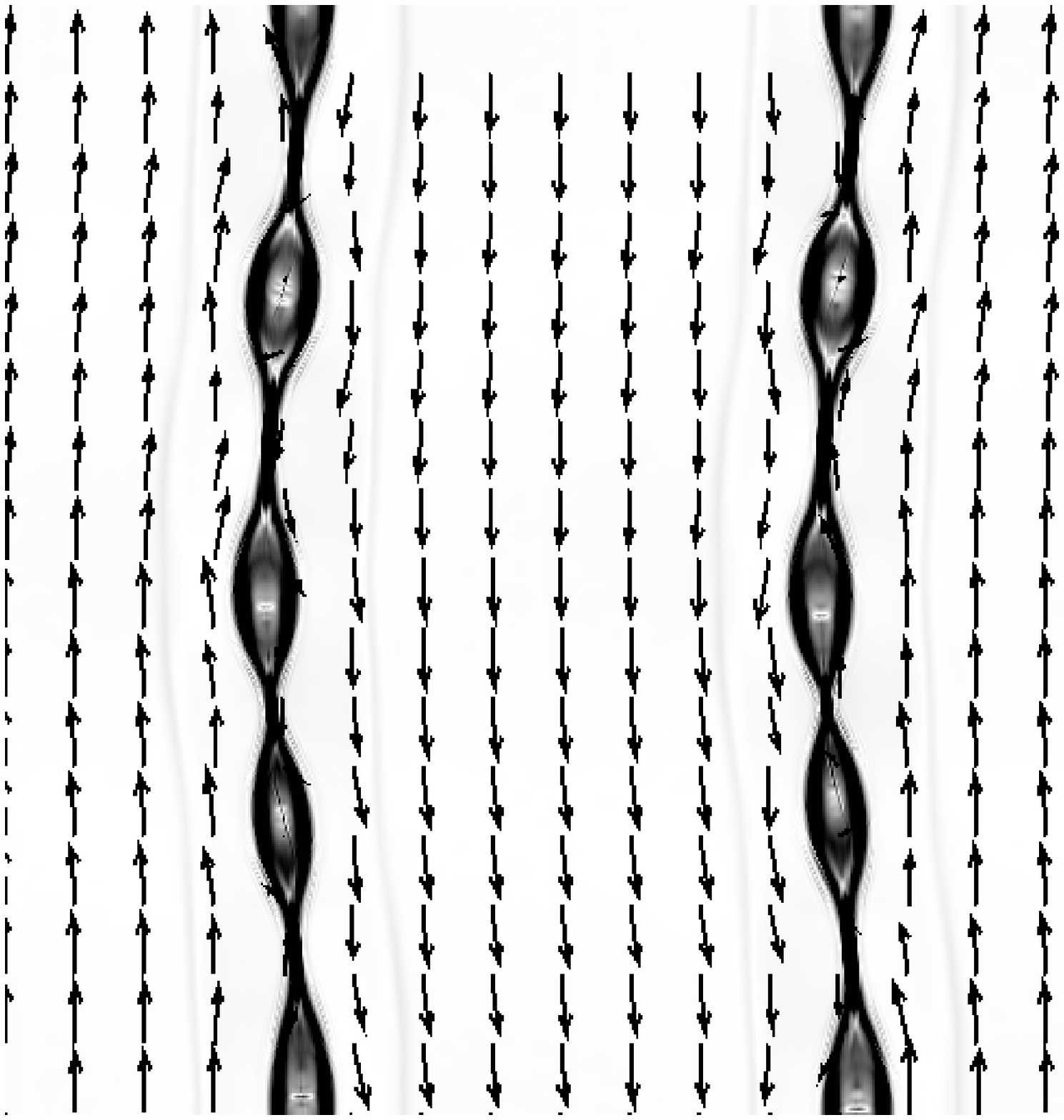}
\includegraphics[width=0.30\columnwidth]{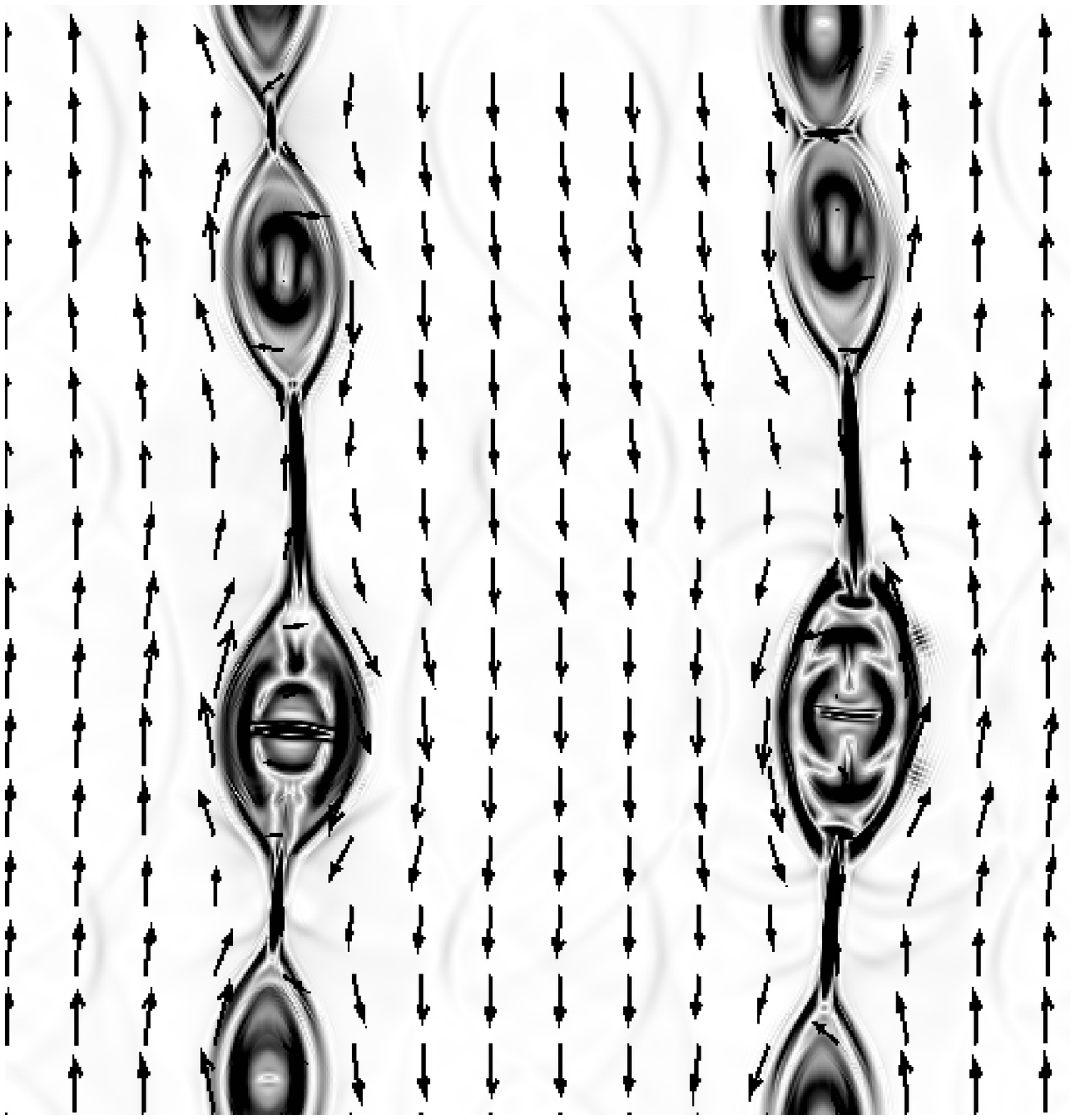}
\includegraphics[width=0.30\columnwidth]{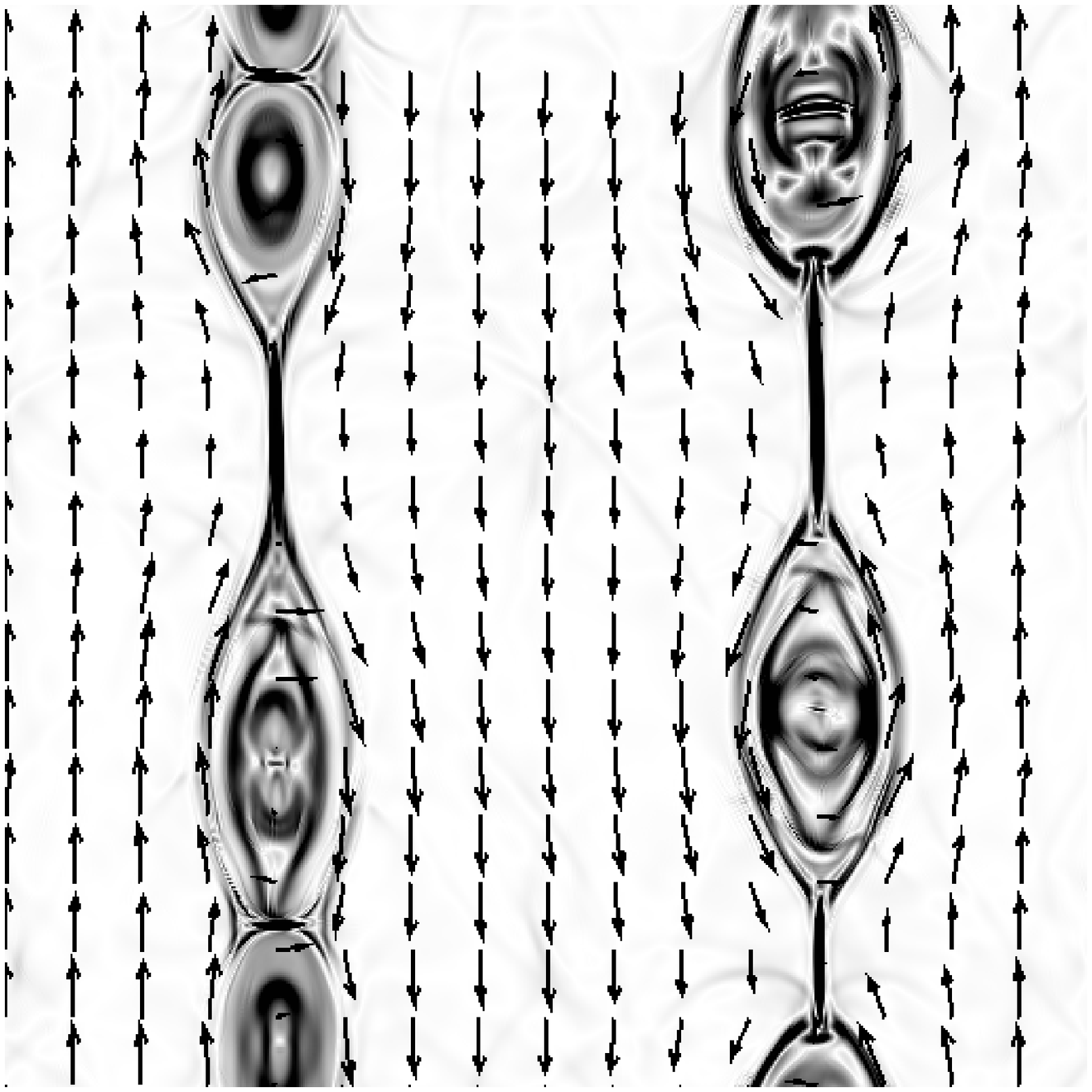}
\includegraphics[width=0.30\columnwidth]{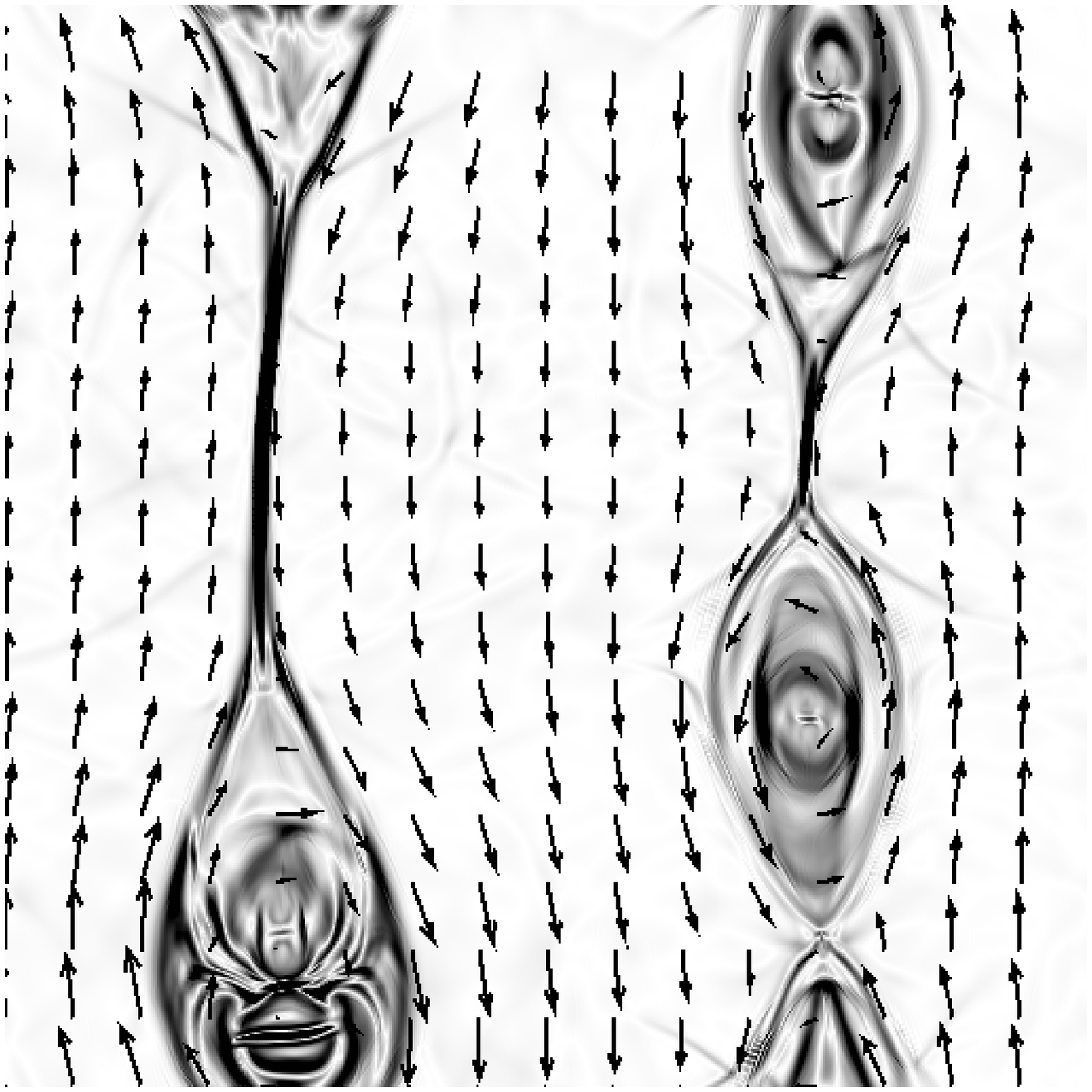}
\includegraphics[width=0.30\columnwidth]{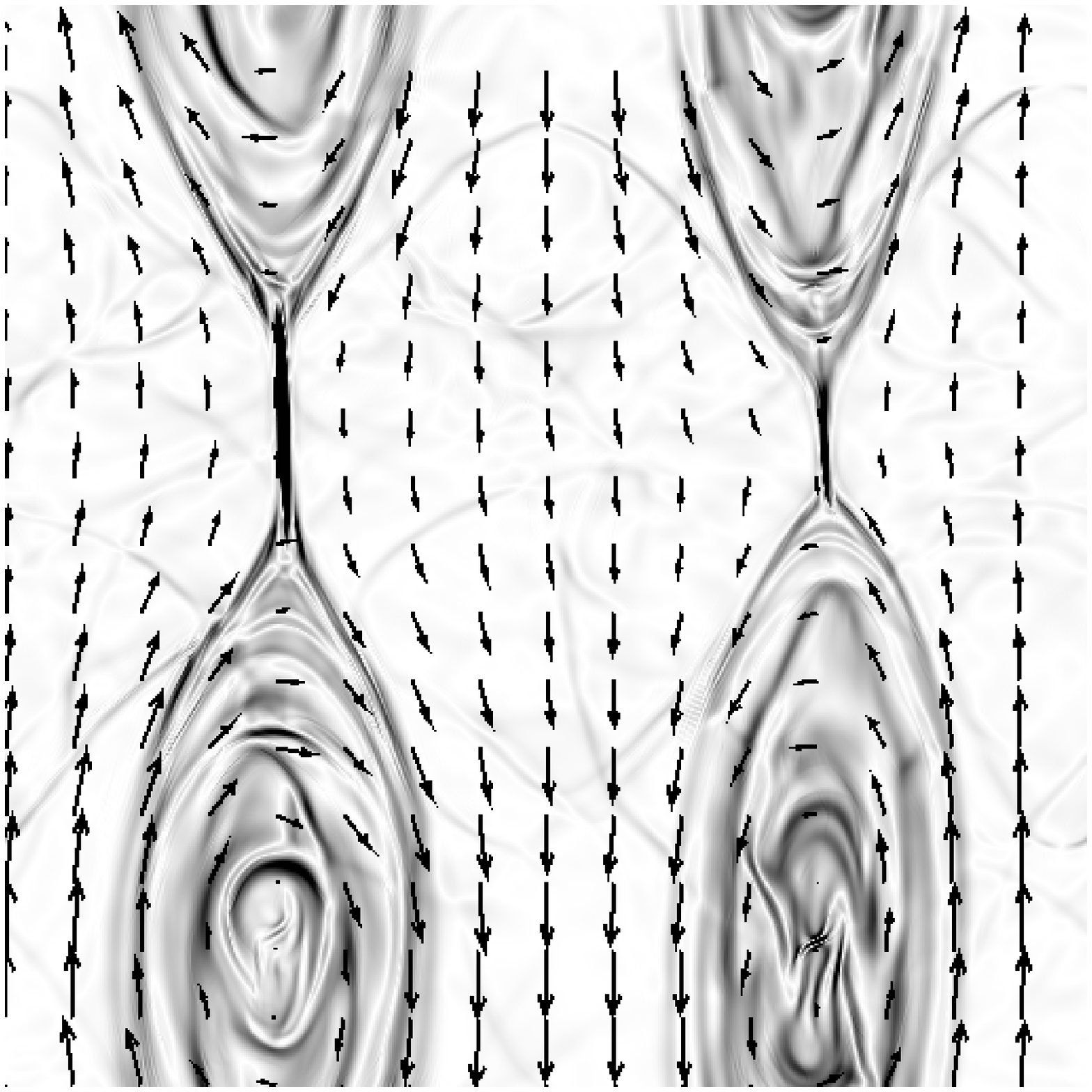}
\includegraphics[width=0.30\columnwidth]{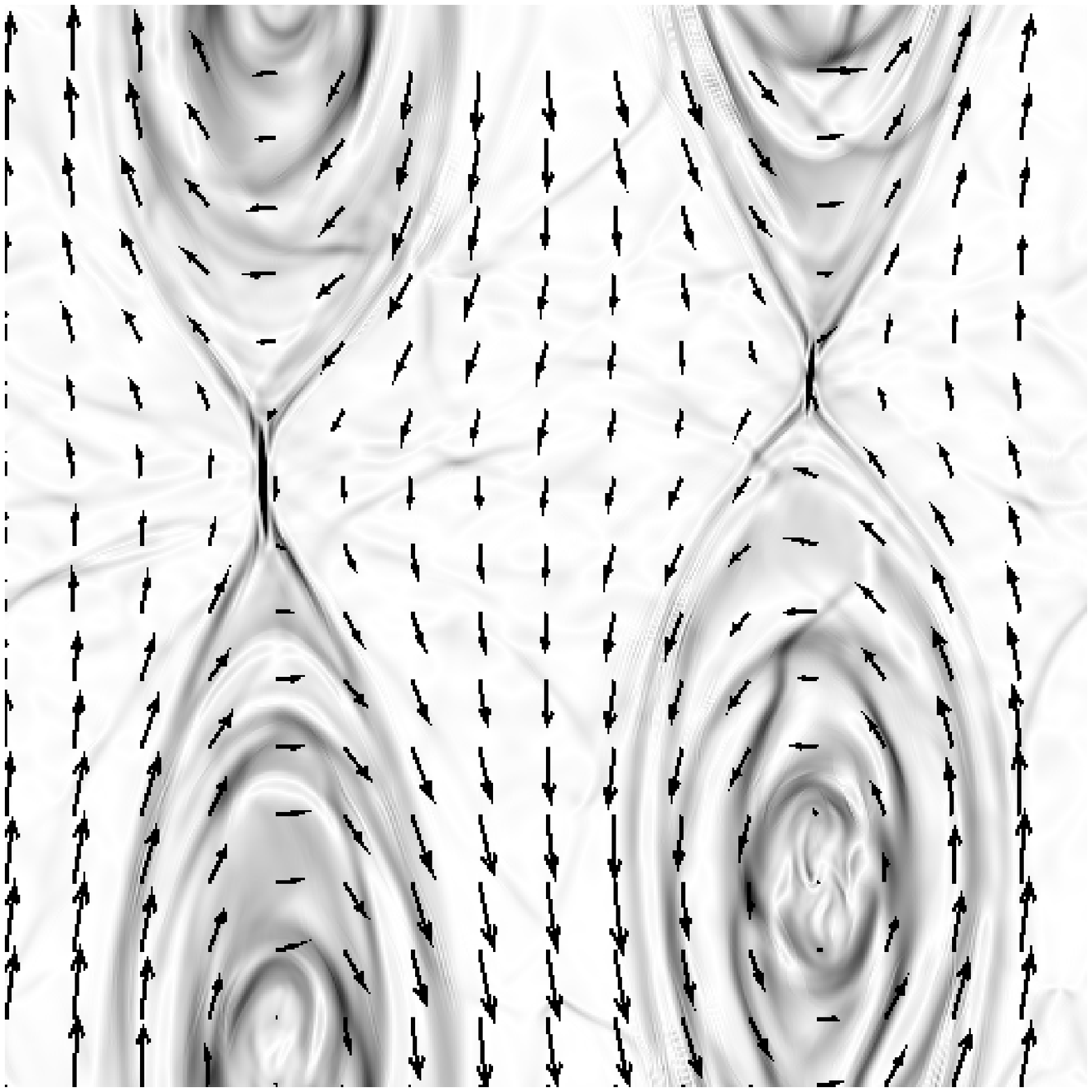}
\caption{Snapshots of tearing instability at $t=0.8, 1.2, 1.6, 2, 4, 6$ with anomalous
resistivity. Grayscale, vectors and resolution as in Fig.~\ref{fig:uniform-resistivity}.}
\label{fig:localized-resistivity}
\end{figure}

\noindent To simulate magnetic reconnection with anomalous resistivity we assume $\eta_1 = 0$, $\eta_2 = 10^{-4}$ and $J_{\indx{crit}} = 100$. The results of a 2D
simulation of the tearing instability, with the initial condition specified at
the beginning of this section are shown in 
Fig.~\ref{fig:localized-resistivity}. We find that in the case of localized
resistivity multiple magnetic islands are formed along each current sheet, and
the reconnection process is apparently more localized than in the uniform
resistivity case. During the evolution smaller islands undergo coalescence to
form bigger ones. Finally, two large magnetic islands remain and the
reconnection tends to form small--size current sheets, resembling 'X--type' reconnection points of
Petschek's reconnection model \cite{petschek}. The reconnection process is
faster in the case of localized resistivity, since the final volume of magnetic
islands is apparently larger than in the uniform resistivity case.
Although magnetic reconnection can be observed in 'ideal MHD' simulations
due to numerical resistivity, utilization of the resistivity module allows
for controlling the speed and properties of the reconnection process in
MHD simulations.
\newpage
\section*{Acknowledgements}
This work was partially supported by Nicolaus Copernicus University through
the grant No. 409--A, Rector's grant No. 516--A, by European Science Foundation within the
ASTROSIM project and by Polish Ministry of Science and Higher Education through
the grants \mbox{92/N--ASTROSIM/2008/0}~and~\mbox{PB 0656/P03D/2004/26.}
%
%

\end{document}